\documentclass[fleqn,twoside,a4paper]{article}
\usepackage{adpsremulate}
  
\usepackage{graphicx}
\usepackage[figuresright]{rotating}
 
\usepackage{amssymb,psfig,times}
\usepackage{url}
\usepackage{natbib}
\usepackage{natbibspacing}
\usepackage{floatflt}

\newcommand{\adspr}{{AdSpR}}
\newcommand{\apss}{{Astroph. and Space Science}}
\newcommand{\apj}{{ApJ}}
\newcommand{\apjs}{{ApJS}}
\newcommand{\apjl}{{ApJ}}
\newcommand{\aj}{{AJ}}
\newcommand{\aap}{{A\&A}}
\newcommand{\aaps}{{A\&AS}}
\newcommand{\nat}{{Nat}}

\newcommand{\prc}{{PhRvC}}
\newcommand{\araa}{{ARA\&A}}

\newcommand{\pasp}{{PASP}}
\newcommand{\npa}{{NuPhA}}

\newcommand{\chandra}{{\it Chandra}}

\newcommand{\rosat}{{\it ROSAT}}

\newcommand{\smm}{{\it SMM}}
\newcommand{\sax}{{\it BeppoSAX}}
\newcommand{\rxte}{{\it RXTE}}
\newcommand{\osse}{{\it OSSE}}

\newcommand{\integr}{{\it INTEGRAL}}

\newcommand{\comptel}{{\it COMPTEL}}
\newcommand{\cgro}{{\it CGRO}}

\newcommand{\casa}{{Cas\,A}}

\newcommand{\gray}{{$\gamma$-ray}}
\newcommand{\grays}{{$\gamma$-rays}}

\newcommand{\tiff}{{$^{44}$Ti}}
\newcommand{\scff}{{$^{44}$Sc}}
\newcommand{\caff}{{$^{44}$Ca}}
\newcommand{\cofs}{{$^{56}$Co}}
\newcommand{\nifs}{{$^{56}$Ni}}
\newcommand{\fefs}{{$^{56}$Fe}}

\newcommand{\arcmin}{{$^{\prime}$}}
\newcommand{\degr}{{\symbol{23}}}
\newcommand{\msun}{{$M_{\odot}$}}
\newcommand{\net}{{$n_{\rm e}t$}}

\newcommand{\fluxunit}{{ph\,cm$^{-2}$s$^{-1}$}}
\newcommand{\kms}{{km\,s$^{-1}$}}

\def\sep{;}

\title{\centerline{
Gamma-Ray Observations of Explosive Nucleosynthesis Products}
}
\author{\centerline{{\Large Jacco Vink
\address{\centerline{SRON National Institute for Space Research,
Sorbonnelaan 2, 3584 CA Utrecht, The Netherlands}}
}}}

\begin{document}

\begin{abstract}
\centerline{\Large \today}
\setlength{\unitlength}{1cm}
\begin{picture}(16,0.0)(0.4,0.2)
%\put(0,0){\line(1,0){16}}
\put(0,0){\line(1,0){17.7}}
\end{picture}
\section*{Abstract}

In this review I discuss the various \gray\ emission lines
that can be expected and, in some cases have been  observed, 
from radioactive explosive nucleosynthesis products. 
The most important \gray\ lines result from the
decay chains of  \nifs,
$^{57}$Ni, and \tiff.
\nifs\ is the prime explosive nucleosynthesis product
of Type Ia supernovae, and its decay determines to a large extent
the Type Ia light curves.
\nifs\ is also a product of core-collapse supernovae, and in fact,
\gray\ line emission from its daughter product,
\cofs, has been detected from SN1987A by
several instruments. 
The early occurrence of this emission was surprising and indicates that
some fraction of \nifs, which is synthesized in the
innermost supernova layers, 
must have mixed with the outermost supernova ejecta.

Special attention is given to the \gray\ line emission of the decay chain
of \tiff\ (\tiff $\rightarrow$\ \scff  $\rightarrow$\ \caff),
which is accompanied by line emission at 68~keV, 78~keV, and 1157~keV.
As the decay time of \tiff\ is $\sim 86$~yr, one expects this line emission
from young supernova remnants. Although the \tiff\ yield 
(typically $10^{-5}-10^{-4}$~\msun) is not very high,
its production is very sensitive to the energetics and asymmetries
of the supernova explosion, and to the mass cut, which defines
the mass of the stellar remnant. This makes \tiff\ an ideal tool
to study the inner layers of the supernova explosion.
This is of particular interest in light of observational evidence for 
asymmetric supernova explosions.

The \gray\ line emission from \tiff\ has so far only been detected
from the supernova remnant \casa. I discuss these detections,
which were made by \comptel\ (the 1157~keV line) and \sax\ (the 68~keV and
78~keV lines), which, combined, give a flux of 
$(2.6\pm0.4\pm0.5)\times10^{-5}$~\fluxunit\ per line,
suggesting a \tiff\ yield of $(1.5\pm1.0)\times10^{-4}$~\msun.
Moreover, I present some preliminary results
of \casa\ observations by \integr, which so far has yielded a $3\sigma$
detection of the 68~keV line with the ISGRI instrument with a flux that 
is consistent with the \sax\ detections. 
Future observations by \integr-ISGRI will be able to constrain the 
continuum flux above 90 keV, as the uncertainty about the continuum shape,
is the main source of 
systematic error for the 68 keV and 78 keV line flux measurements.
Moreover, with the \integr-SPI instrument
it will be possible to measure or constrain the line broadening of
the 1157~keV line.  A preliminary analysis of the available data indicates
that narrow line emission (i.e. $\Delta v < 1000$~\kms) can be almost
excluded at the 2$\sigma$ level, for an assumed  line flux of 
$1.9\times10^{-5}$~\fluxunit.

{\it PACS:} 26.30.+k\sep 95.85.Pw\sep 97.60.Bw\sep 98.38.Mz\\
{\it Keywords:} Nucleosynthesis\sep  Supernovae\sep  Supernova remnants\sep 
Gamma-rays\sep  X-rays\sep  individual (Cas A)\sep  individual (SN 1987A)\\
\begin{picture}(16,0.0)(0.0,0.2)
\put(0,0){\line(1,0){17.7}}
\end{picture}
\end{abstract}

\maketitle

\section{Introduction}

Nature's heaviest elements are the products of
explosive nucleosynthesis in supernovae.
A large fraction of the explosive nucleosynthesis products begin as 
radioactive elements. This is in particular the case for Type Ia supernovae,
which, according to the standard scenario, are the result of
thermo-nuclear disruptions of carbon-oxygen white dwarfs with a mass close
to the Chandrasekhar limit. The energy for the disruption comes
from the thermo-nuclear burning of C and O into $\sim0.6$~\msun\  of
\nifs, a radioactive element that is the source
of \fefs, the most abundant iron isotope \citep[see e.g.][]{arnett96}.

For core-collapse supernovae, explosive nucleosynthesis is not the
driver behind the explosion (the energy comes from the gravitational
collapse of the stellar core). The hot, inner regions of
the supernova typically produce 0.1~\msun\ of \nifs, but with a
larger variation in yield from one core collapse event to another.

The most telltale signs of the synthesis of radioactive material
are the lengths and shapes of supernova light curves. 
Although the expanding plasma cools adiabatically, radio-activity
heats the plasma making the supernova visible for an extended period.
%The energy deposition comes from the \gray\ radiation from the radioactive
%elements and from the thermalization, and annihilation of the positrons
%produced by $\beta^+$-decay.
The amount of radioactive material can be estimated from modeling the
light curves, taking into account radiative transfer in the expanding
atmosphere, the possible escape of positrons, and mixing 
\citep{milne01,fransson02}.

The \gray\ emission from the radioactive elements
can in principle be observed directly. Although %especially
in the early phases of the supernova the atmosphere is opaque to \grays,
at later phases \gray\ emission should emerge. In fact the
phase at which \grays\ emerge is an important diagnostic tool in itself
\citep{hoeflich98}. 
The spectral information can be used to measure the velocity
distribution of the radioactive material.
The goal of \gray\ observations of fresh 
radioactive material is to provide new insights into the yields,
% and into the explosion velocities and asymmetries of explosive nucleosynthesis products.
expansion  velocities and asymmetries of the explosive 
nucleosynthesis products.
Moreover, the line emission from the longer lived radioactive element \tiff\
is a tracer of recent supernova activity.\footnote{ 
Another long lived element that traces recent stellar nucleosynthesis 
is $^{26}Al$\ at 1809 keV, with an decay time of $1.1\times10^6$~yr.
However, the most likely dominant source of $^{26}Al$\ is not supernovae,
but the winds of massive stars \citep{prantzos04}.}

For Type Ia supernovae \gray\ observation can be used to
distinguish between various explosion
models, such as ``delayed detonation'' and 
He-detonation of sub-Chandrasekhar mass white dwarfs \citep{woosley94}. 
This is of particular interest
for cosmology, as Type Ia supernovae are used as standard candles 
\citep[e.g.][]{riess04}, but the variation in absolute peak brightness
and its correlation with light curve properties \citep{phillips93}, 
which is used to calibrate the absolute brightness of the supernovae,
is not well understood. 
The brightness variation is likely to reflect a variation in \nifs\ yield,
but the cause of this variation is still under debate 
\citep[e.g.][]{mazzali01,roepke04}.

\begin{figure}
\centerline{
\psfig{figure=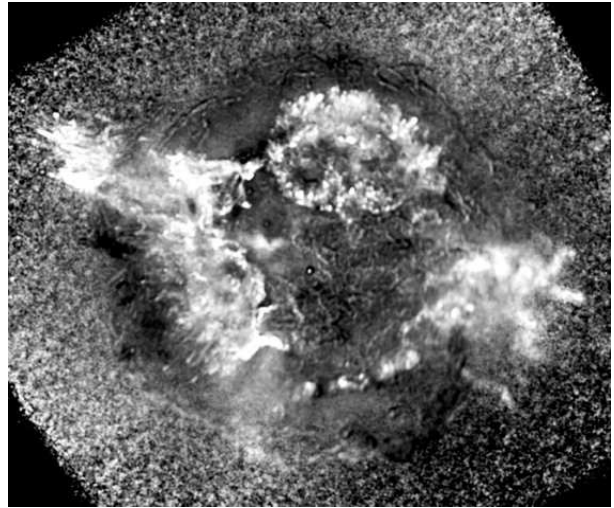,width=8cm}}
\caption{Image based on a very deep, 1 Ms, \chandra\ observation of \casa,
which has been processed to bring out the jet/counter-jet structure. 
%\citep{vink04a,hwang04}.
(Credit: NASA/CXC/GSFC/U.Hwang et al.)\label{jet}}
\end{figure}

For core-collapse supernovae the prime interest is to understand
the nature of the explosion mechanism. This has become more important now
that gamma-ray bursts appear to be 
related to
%very energetic 
core-collapse supernovae, which challenges our view of the core-collapse
explosion mechanism in general. 
There are two main hypotheses about the likely bipolar nature
of gamma-ray bursts,
the collapsar model \citep{macfadyen01}, or, alternatively, 
jet-formation due to magneto-rotational instabilities
\citep{akayima03}.
In particular the latter hypothesis may, to some extent, be relevant
for all collapse explosions. A case in point is the discovery
of a jet/counter jet system in the supernova remnant 
\casa\ \citep[Fig.~\ref{jet},][]{vink04a,hwang04,hines04},
and the non-spherical expansion of its bright Si-rich shell
\citep{markert83,willingale02}.
On the other hand, recent
simulations of more conventional core-collapse supernovae
indicate a highly turbulent nature of the inner regions,
which may give rise to \nifs-rich high-velocity material.
%\citep{kifonidis03}.
This high velocity material may finally be ejected with
a high velocity,
if the star has a low envelope mass,
e.g. due to stellar wind mass losses,
as is thought to be the case for Type Ib/c supernovae \citep{kifonidis03}.
This may be the cause of the Fe-rich ``bullets'' in  
\casa\ that have overtaken the Si-rich shell of ejecta \citep{hwang03}.
\casa, a likely Type Ib remnant,  
will feature prominently here, as it is the only confirmed source
of \tiff\ \gray\ emission, 
and an important target of the \integr\ \gray\ mission.

%I will discuss the past observations of \grays\ of explosive nucleosynthesis
%products, and present some preliminary results from  a deep observation
%of \casa\ by \integr.

\begin{table*}
\begin{center}
\parskip 0mm
\caption{
Decay chains and \gray\ signatures 
  of shortlived radioactive products from explosive SN nucleosynthesis.
\label{decay}}
{%\footnotesize %\small
\begin{tabular}{lllrlr}
\noalign{\smallskip}\hline\noalign{\smallskip}
&         &  &  Decay time & Process &Lines (keV)\\
\noalign{\smallskip}
\hline
\noalign{\smallskip}
$^{56}$Ni $\rightarrow$ &                         &         &8.8 d &
EC& 158, 812 \\
                        &$^{56}$Co $\rightarrow$  &$^{56}$Fe& 111.3 d& 
EC, $e^+$ (19\%) & 847, 1238\\ %, 2599, 3250\\
\noalign{\smallskip}
$^{57}$Ni $\rightarrow$ &                         &         & 52 hr & 
EC &1370\\
                        &$^{57}$Co $\rightarrow$ & $^{57}$Fe & 390 d & 
EC & 122\\
\noalign{\smallskip}
$^{44}$Ti $\rightarrow$ & & &86.0 yr & EC &67.9, 78.4\\
  &$^{44}$Sc $\rightarrow$ &  $^{44}$Ca& 5.7 hr & 
 $e^+$, EC (1\%) &1157\\
\noalign{\smallskip}
\hline
\end{tabular}
}
\end{center}
\end{table*}

\section{Gamma-ray lines from supernovae}% nucleosynthesis products}
% in SN 1987A}
Table~\ref{decay} list the most important radioactive elements
produced by explosive nucleosynthesis \citep[for more general
reviews on \gray\ line emission, see][]{diehl98,prantzos04}. 
In the case of electron capture, additional lines in the X-ray band
will arise from characteristic de-excitations of the atomic shells of a few key
isotopes \citep{leising01}.
%In the case of electron capture, the decay will also be accompanied
%by electron de-excitations observable in the X-ray band \citep{leising01}.

The most abundant radioactive nucleosynthesis product,
\nifs, is the product of Si-burning and $\alpha$-rich freeze-out,
whereas \tiff\ is almost exclusively the product of  $\alpha$-rich freeze-out in core-collapse
supernovae \citep{arnett96,the98}, although sub-Chandrasekhar mass Type Ia
supernovae may be an additional source of \tiff\ \citep{woosley94}.
Alpha-rich freeze-out occurs
when material that is initially in nuclear statistical equilibrium
cools adiabatically in the presence of an $\alpha$\ particle excess
caused by the triple-alpha reaction bottleneck.
This makes the \tiff\ yield very sensitive to the explosion
energy and asymmetries, as faster expansion causes a more rapid freeze out.
The yield is also very sensitive to the place of the mass cut\footnote{
The mass cut defines the boundary of material that will be ejected, 
versus what will fall back on the stellar remnant (neutron star or black hole).
},
as \tiff\ is synthesized in the deepest supernova layers, 
close to the mass cut.

 Finally, \grays\ observed from the inner Galaxy from 
electron-positron annihilation, consisting of 511~keV line emission,
and three photon continuum decay of positronium,
may also find its origin in explosive nucleosynthesis.
The reason is that the $\beta$-decays of \cofs\ and \scff\ produce
positrons.
Due to its long lifetime
almost all positrons from \scff\ 
will escape the supernova, but this is less clear for
\cofs, as its decay takes place during the earlier phases
of the supernova, when the density is high \citep{colgate70,chan93,milne99}.
However, more \cofs\ positrons will escape if the expansion is fast
\citep{casse04,vink04a}. This led \citet{casse04} to the suggestion
that a hypernova in the Galactic center region
may be responsible for the 511~keV electron-positron
line emission from that region 
\citep[][Kn\"odlseder et al. 2005, in preparation]{kinzer01,knoedlseder03}

\begin{figure}
\centerline{
\psfig{figure=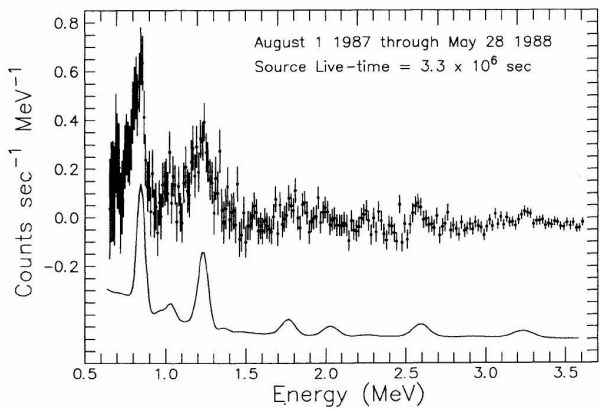,width=8cm}}
\caption{The \gray\ spectrum of SN 1987A, as accumulated
by the \smm-GRS over the period from August 1987 to May 1988
\citep{leising90}. 
Note that apart from the prominent 847~keV and 1238~keV lines,
there is also evidence for the weaker 2599~keV and 3250~keV lines
from \cofs. (Courtesy M. Leising)}\label{figleising}
\end{figure}

\subsection{Detections of \cofs\ lines from SN 1987A.}
The importance of future \gray\ observations of supernovae can be easily
justified by the impact of the detection of \gray\ lines
from SN 1987A.
The 847~keV and 1238~keV \cofs\ lines were detected by a number
of balloon experiments \citep{cook88,mahoney88,sandie88,teegarden89}, 
and the gamma-ray spectrometer (GRS)
on board the Solar Maximum Mission (\smm) satellite 
\citep[][see Fig.~\ref{figleising}]{matz88a,leising90}.
The observed fluxes were typically in the range 
$(0.5-1)\times 10^{-3}$~\fluxunit, indicating that only a few percent
of the total $\sim 0.075$~\msun\ of \cofs\ was exposed.
The detection of 122~keV line emission from $^{57}$Co by
the Oriented
Scintillation Spectrometer Experiment (\osse) on board the Compton
Gamma Ray Observatory (\cgro) was later reported by \citet{kurfess92}.

Although the emergence of \cofs\ \gray\ lines had been predicted
\citep{clayton69}, it was a surprise that the lines became observable
as early as 160 days after the explosion \citep{matz88a}.
\cofs\ was expected to be buried too deep into the ejecta to be
observable, but the early detection
indicates that at least some of the \cofs\ must have mixed out
to larger radii. In velocity coordinates this means
that $\sim $5\% of iron-group rich material must have mixed
out to velocities of $\sim 3000$~\kms\ 
\citep[][and reference therein]{arnett89,leising90}, 
while most of the iron-group material is thought to have
a velocity less than 1000~\kms.
The cause for this mixing may be Rayleigh-Taylor instabilities,
and, in addition, the expansion of  \nifs/\cofs-rich bubbles 
caused by radioactive heating \citep[e.g.][]{basko94,kifonidis03}.

Models for \gray\ emission that include ejecta mixing predict that
the \gray\ lines should be blue-shifted, as only the \grays\ from
the expanding sphere facing us should be observable \citep{pinto88b}.
However, the high spectral resolution  observations of SN~1987A showed
that the 847~keV and 1238~keV lines were broadened by $\sim 3000$\kms\ (FWHM)
and {\em red}shifted by $\sim 400$\kms\ 
\citep{sandie88,teegarden89,tueller90}.
This discrepancy is still not understood, but it clearly points
to an asymmetric explosion. The \gray\ emission from SN 1987A thus adds
another piece to the puzzle of understanding 
the nature of core-collapse supernovae.

\subsection{Gamma-ray observations of Type Ia supernovae}

 As explained above, nucleosynthesis in Type Ia supernovae,
is the cause for the explosion, and not merely a by-product of the explosion,
as for core collapse supernovae.
Gamma-ray observations of Type Ia therefore promise to give new
insights into the light curve variation among Type Ia supernovae.
Unfortunately, no solid detections of \nifs\ and \cofs\ line emission
have been made, 
as there has been a lack of sufficiently nearby Type Ia events.
The best candidate for \gray\ observations was SN 1991T, a relatively
bright Type I supernova \citep{phillips93} at a distance of $\sim 13.5$~Mpc.

Both \osse\ and the Compton Telescope (\comptel) on board \cgro\
observed this supernova remnant. An initial analysis of the \comptel\ data
and the \osse\ data resulted in upper limits on \cofs\ lines
fluxes of $3-4\times10^{-5}$~\fluxunit\
\citep[resp.][]{lichti94,leising95}. However, a reanalysis
of the \comptel\ observation showed evidence for
line emission at the $\sim3\sigma$\ level \citep{morris97}.
Although it is uncertain whether the \comptel\ detection holds up, %is real,
the predicted \cofs\ line fluxes for SN 1991T 
were close to the \comptel\ and \osse\ upper limits. 
This gives some hope that \integr\ will be able to detect a not too distant
Type Ia supernova in the near future.

%At the time of the explosion, no satellite \gray\ satellites were in orbit,
%but several balloon experiments

\begin{figure}
\centerline{
\psfig{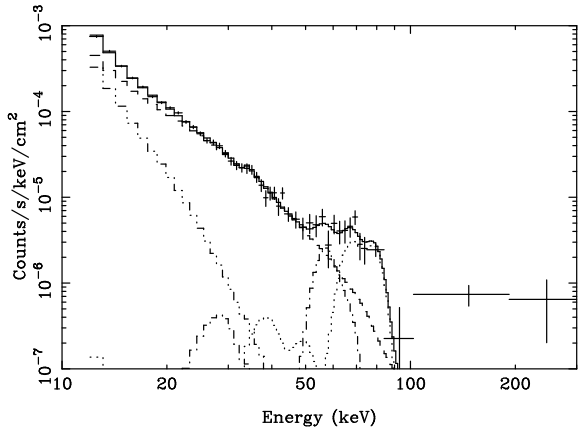}}
\caption{\sax-PDS spectrum of \casa\ \citep{vink01a,vink03a}. 
%The raw count rate spectrum
%has been divided by the effective area, in order to obtain aproximate
%flux densities.
The continuum model consists of a thermal (dash-tripled dotted line)
and non-thermal bremsstrahlung model (dashed), 
the latter based on electron acceleration
by lower hybrid wave \citep{laming01b}.
The 67.9~keV and 78.4~keV line contributions are indicated
jointly by a dotted line. In addition, a potential
instrumental background line of  Ta-K$\alpha$\ is indicated
by a dash-dotted line.
The observed  count-rate spectrum has been
divided by the instruments effective area in order to obtain an approximate 
flux density estimate. 
\label{pdsspec}}
\end{figure}
\subsection{\tiff\ detections by COMPTEL \& BeppoSAX}
\label{secCasA}
One of the highlights of the \comptel\
experiment is the detection of
the 1157 keV line associated with the decay of \tiff\ \citep{iyudin94}.
\casa\ is the youngest known Galactic supernova remnant,
and is probably the result of a Type Ib or IIb supernova
\citep[][for a review]{vink04a}.
So of all remnants \casa\ was the most likely source of \tiff.
On the other hand, \casa\ exploded around 1672 and given the expected
\tiff\ yields of around $10^{-5}$~\msun\ \citep[e.g.][]{ww95}, 
the 1157 keV line flux should have been below the \comptel\ detection limit.
Surprisingly, \comptel\ detected the 1157 keV line emission, with the latest,
most reliable flux estimate based on \comptel\ observations being
$(3.4\pm0.9)\times 10^{-5}$~\fluxunit\ \citep{schoenfelder00}.
%$(7.0\pm1.7)\times 10^{-5}$~\fluxunit, indicating a \tiff\ yield
%of $(1.4 - 3.2)\times10^{-4}$ \msun\ \citep{iyudin94}.

The \tiff\ decay chain also produces line emission
at 67.9~keV and 78.4~keV. Several observation with \osse, \rxte-HEXTE, and
\sax-PDS were made in order to confirm the high \tiff\ flux, but
without success \citep{the96,rothschild98,vink00b}.
The most stringent flux upper limit ($<4\times10^{-5}$~\fluxunit), 
obtained with the \sax-PDS,
was in fact below the original \comptel\ flux measurement \citep{vink00b}.

This discrepancy was finally solved with a 500~ks observation of
\casa\ by \sax\, which resulted in the first detection
of the 67.9~keV and 78.4~keV lines \citep[][see Fig.~\ref{pdsspec}]{vink01a}.
%The flux appeared to be lower than indicated by the first \comptel.
A complication for measuring the line fluxes of the low energy lines is the
presence of continuum emission from \casa.
The nature of the continuum emission is unknown,
and interesting in its own right. It could be
either due to 
synchrotron radiation from electrons with energies in the TeV range 
\citep{allen97,reynolds98}, or it could arise from bremsstrahlung by 
supra-thermal electrons,
which have been accelerated by internal shocks \citep{laming01a,laming01b},
or by Fermi acceleration at the forward shock \citep{asvarov90}.
Both the synchrotron and the non-thermal bremsstrahlung model
by Laming predict a gradual steepening at high energies.
Unfortunately there are no reliable continuum flux measurements
at energies just above the 78.4~keV line.
As a result, assuming that the continuum is a simple power-law spectrum 
gives a line flux for each line of
($1.9\pm0.4)\times10^{-5}$~\fluxunit\
(68\% confidence range), whereas the gradual
steepening of synchrotron and specific non-thermal bremsstrahlung
models results in a  line flux of
($3.2\pm0.3)\times10^{-5}$~\fluxunit\ \citep{vink01a,vink03a}.
Within the errors both values are consistent with the latest
\comptel\ results \citep{schoenfelder00}.
We can average the \comptel\ and \sax\ results
to obtain a \tiff\ line flux of
$(2.6\pm0.4\pm0.5)\times10^{-5}$~\fluxunit\ per line, 
where the first error is the
1$\sigma$ statistical error and the second error is the systematic error,
due to uncertainty in the continuum modeling.\footnote{
Note that \citet{vink01a} list  90\% errors, whereas for
weighing the results and assessing the total
statistical error 1$\sigma$\ (68\%) errors are used \citep[c.f.][]{vink03a}.}

%Both values, are consistent with the latest flux measurements
%based on \comptel\ observations by \citet{schoenfelder00}.
%, which, after accumulation of more
%data and refining the analysis methods, {\bf
%\citet{schoenfelder00} reported a flux of
%$(3.4\pm0.9)\times10^{-5}$~\fluxunit.}

\subsection{The \tiff\ yield of \casa}
The detection of \tiff\ in \casa\ led to a renewed interest in measuring
the life time of \tiff, which was highly uncertain, 
with published life times varying between 67~yr and 96~yr \citep{diehl98}. 
In 1998, however, accurate measurements by three independent groups
give a life time of 85 yr \citep{ahmad98,goerres98,norman98}.
Meanwhile more measurements have been published in agreement
with these results \citep{wietfeldt99,hashimoto01}. 
The error weighted average of these results gives a life
time of $86.0\pm0.5$~yr.

The distance to \casa\ is $3.4^{+0.3}_{-0.1}$~kpc \citep{reed95},
and the most recent estimate of the supernova event is 
A.D. $1671.3\pm0.9$ \citep{thorstensen01}.\footnote{
This is close to the putative data of A.D. 1680 of the detection of
a putative supernova by Flamsteed \citep{ashworth80}, but 
the connection of Flamsteed's observation with the supernova is
debatable, and rejected by \citet{stephenson02}.}
The \gray\ emission implies an initial \tiff\ mass of 
$M_0($\tiff$) = (1.6\pm 0.3\pm0.3)\times10^{-4}$~\msun.
This is comparable to the \tiff\ yield from SN 1987A, 
which, based on the light curve, is estimated to be
$M_0($\tiff$) = (1.5\pm1.0)\times10^{-4}$~\msun\
\citep{motizuki04}. An estimate based
on the forbidden Fe~I and Fe~II infrared line emission suggests
that the \tiff\ yield was slightly lower than for \casa, 
$M_0 \lesssim 1.1\times 10^{4}$~\msun\ \citep{lundqvist01}.

\begin{figure}[t]
\centerline{
\psfig{figure=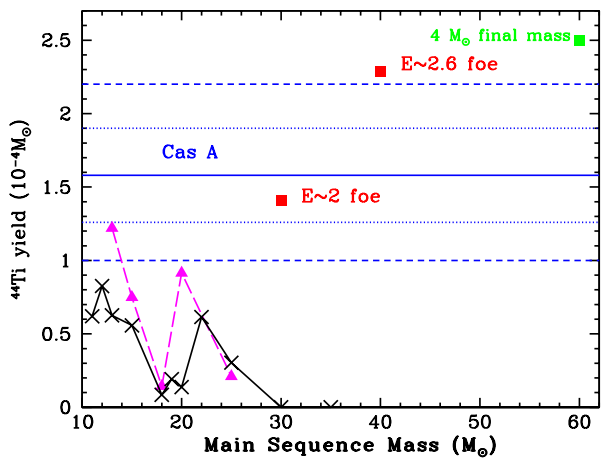,angle=-90,width=\columnwidth}}
\caption{\tiff\ yield as a function of main sequence mass.
The standard models are shown as crosses \citep[based on][]{ww95}
and triangles \citep{nomoto97}. Also shown are the 
models by \citet{ww95} for which the explosion energy was increased
(1 foe $= 10^{51}$~erg) and the Type Ibc model of
a single massive star that exploded with a final mass of 4~\msun\ 
\citep{wlw93}.
\label{yield}}
\end{figure}

\subsection{Implications of the \tiff\ yield}
\label{implications}
A comparison of the \tiff\ yield of \casa\ with model calculations
(Fig.\ref{yield}) shows that nucleosynthesis models \citep{ww95,nomoto97}
under-predict the observed yield. 
Taken at face value this implies that the supernova event must have
been a more energetic event (e.g. $2\times10^{51}$~erg instead of the
canonical  $10^{51}$~erg), an asymmetric supernova \citep{nagataki98},
or the result of the explosion of a star that lost most of its mass
\citep{wlw95}.
As for the effects of asymmetry, \citet{maeda03} consider various
bipolar supernova models, which indeed give a higher \tiff\ yield.
However, the higher \tiff\ yields of their models seem to be
largely caused by the high kinetic energy, $\sim10^{52}$~erg, of the 
explosion models rather than due to the bipolar nature of the explosions,
as a comparison of the bipolar and spherical models 
(resp. models 40A and 40SHa) shows. On the other hand,
the bipolar models have a higher \tiff/\nifs\ ratio, which is in
better agreement with the \tiff\ and \nifs\ estimates of
SN 1987A.

Both mass loss and a higher explosion energy result in a reduced fall
back of ejecta on the proto-neutron star/black hole, since both affect
the final velocity of the ejecta.
It can be argued that for \casa\ all those effects play a role.
After all, \casa\ was probably a Type Ib supernova, and the ejecta mass
as measured in  X-rays is quite low, in the range 2-4\msun. Furthermore,
the kinetic energy, as derived from the supernova remnant's kinematics 
\citep{vink98a,delaney02} 
has been estimated to be $\sim 2\times10^{51}$~erg 
\citep{laming03}.
Finally, there is ample evidence for asymmetric
ejecta expansion \citep{hwang03,vink04a,hwang04}.

The \tiff\ yield, therefore, strengthens our current ideas about
\casa, but one should keep in mind that the predicted \tiff\ yields
are uncertain. Most simulations are based on one-dimensional models, 
in which artificially energy or momentum is deposited at the base of the 
ejecta. Multidimensional models of core-collapse supernovae
show that the ejecta kinematics is highly turbulent, which is likely to have 
an  effect on the \tiff\ yield.
Moreover, there is still uncertainty whether
magnetic fields and rotation are important for the explosion mechanism  
\citep{akiyama03}.

It is worthwhile to note that the high fluxes
associated with \tiff\ decay from \casa, might also be the result of
a change in effective life-time of \tiff,
as pointed out by \citet{motizuki99,motizuki04}.
The reason is that \tiff\ decays through electron capture, which in most
cases involves the capture of one of the inner most, K-shell, electrons.
If the \tiff\ is sufficiently
ionized, i.e. one or both K-shell electrons are absent,
the electron capture rate drops significantly.
However, to have an effect
the ionization should have taken place early in the life of the remnant, 
in order to inhibit the decay of a substantial amount of \tiff.
This is unlikely to be the case for \casa, given the fact
that X-ray spectroscopy indicates that 
both Ca and Fe, which have similar ionization
cross sections, are still ionizing\footnote{
Even when instantly heated it takes some time before an atom is ionized up to
equilibrium ionization. 
This is governed by the product of electron
density and time, \net. A hot plasma is roughly in equilibrium ionization
if \net$\sim10^{12}$ cm$^{-3}$s, whereas in \casa\ it is typically
\net$\sim10^{11}$ cm$^{-3}$s.}
and are presently in the He-like state \citep[e.g.][]{vink96}. 
This means that 
the K-shell is fully populated, and only a modest effect of $\sim $10\%
on the life time is to be expected. 
Moreover, a substantial fraction of \tiff\ may
not yet have been heated by the reverse shock. 
Finding this out is one of the goals
of \integr\ observations of \casa.
More detailed hydrodynamical calculations 
also indicate that the ionization history has only a
minor effect on the \tiff\ line flux 
\citep{laming01c}.

\subsection{Potential other sources of \tiff\ \gray\ emission}

The Galactic supernova rate is thought to be approximately 2 per century.
However, the latest observed supernova was observed in A.D. 1604 
\citep[Kepler's supernova][]{stephenson02}, and the youngest remnant known,
\casa, was also a 17th century event.
It is unlikely\footnote{The chance that no supernova
occurred in  the Galaxy since 1672, 
given a rate of 2 per century, is 0.17\%.}
that these two supernovae were truly the last, so other
supernovae must have been
too dim to be noted, and their remnants may still escape our attention.
The reason could be that Galactic supernovae have been heavily obscured,
and their remnants are faint because they evolve in low density regions.

The \gray\ emission from \tiff, however, does not depend on
the environment of the supernova, and \grays\ can uninhibitedly
penetrate  the interstellar
medium. Gamma-ray observations of \tiff\ line emission
is therefore a possible method for discovering young remnants.

A search for \tiff\ in the Galactic plane was made with \comptel,
initially without finding
any obvious candidate sources \citep{dupraz97,iyudin99},
but later a potential \tiff\ source,  GRO J0852-4642, 
was discovered
close to the Vela supernova remnant \citep{iyudin98}.
The Vela supernova remnant is too old to contain a significant
amount of \tiff, but \rosat\ observations of the Vela region
showed the presence of a supernova remnant,
RX~J0852.0-4622/G266.2-1.2, in projection
to the Vela supernova remnant \citep{aschenbach98},
which is often affectionately referred to as ``Vela Jr''.

Although, the \comptel\ observation triggered the discovery of the
new remnant, it is not clear whether it is truly a source of \tiff.
First of all, the X-ray absorption is relatively high \citep{slane01a}, 
indicating
a distance beyond the Vela (``Sr'') remnant, but it is also rather
extended, 2\degr. Together this suggests a remnant with an age
of more than 1000~yr, which
is hard to reconcile with the observed 1157~keV line flux
of $(3.8\pm0.7)\times10^{-5}$~\fluxunit\ \citep{aschenbach99,slane01a}.
Also the detection itself, although formally
at the $5\sigma$ level is less secure than
the detection of \tiff\ in \casa\ \citep{schoenfelder00}.
Nevertheless, Vela Jr is an interesting object and 
an important \integr\ target \citep{vonkienlin04}.

%\subsection{K-shells vacancies}

\section{INTEGRAL}
The International Gamma-Ray Astrophysics Laboratory (\integr)
is ESA's \gray\ imaging and spectroscopy mission and
was launched in October 2002 \citep{cwinkler03}. 
The two main \gray\ instruments are the spectrometer SPI \citep{vedrenne03} 
and the imager IBIS \citep{ubertini03}, 
both of which use coded masks to obtain positional information.

IBIS has a fully coded field of view of 5\degr$\times$5\degr, and an
angular resolution of 3\arcmin. Its principal detector plane
instrument is ISGRI \citep{lebrun03}, consisting of 128$\times$128 CdTe 
detectors. SPI, on the other hand, uses 19 Ge detectors\footnote{
Unfortunately, one of the Ge detectors failed in  December 2003, and another
in July 2004, decreasing the sensitivity of the instrument by 10\%.},
giving
a limited spatial resolution of  2.5\degr\ (FHWM), but a 
spectral resolution of $\Delta E = 3$~keV  over a wide energy range 
from 20~keV - 8 MeV. This results in a very high line resolving power,
in particular for MeV energies.

SPI is the main instrument for \gray-line emission, and has already
obtained results on the 1809~keV $^{26}$Al line emission 
from the Galactic plane \citep{diehl04}, and the 511 keV annihilation
line from the Galactic Center \citep{jean03,knoedlseder03}.
Moreover, a comprehensive study of the available data puts new constraints
on the 511~keV emission from the Galactic plane \citep{teegarden04}.

One of the main mission goals of \integr\ is to detect and measure
\tiff\ emission from young known, or still to be discovered, 
supernova remnants.
There are several projects to observe \tiff\ line emission.
The pointed observations concern the (potential) \tiff\ sources
``Vela Jr'' (section \label{velajr}), SN 1987A, Tycho
and \casa. For preliminary \integr\ 
results on ``Vela Jr'' and supernova remnants
searches in the \tiff\ lines, see \citet{vonkienlin04} and \citet{renaud04}.

\begin{figure*}
\vbox{
\centerline{
\psfig{figure=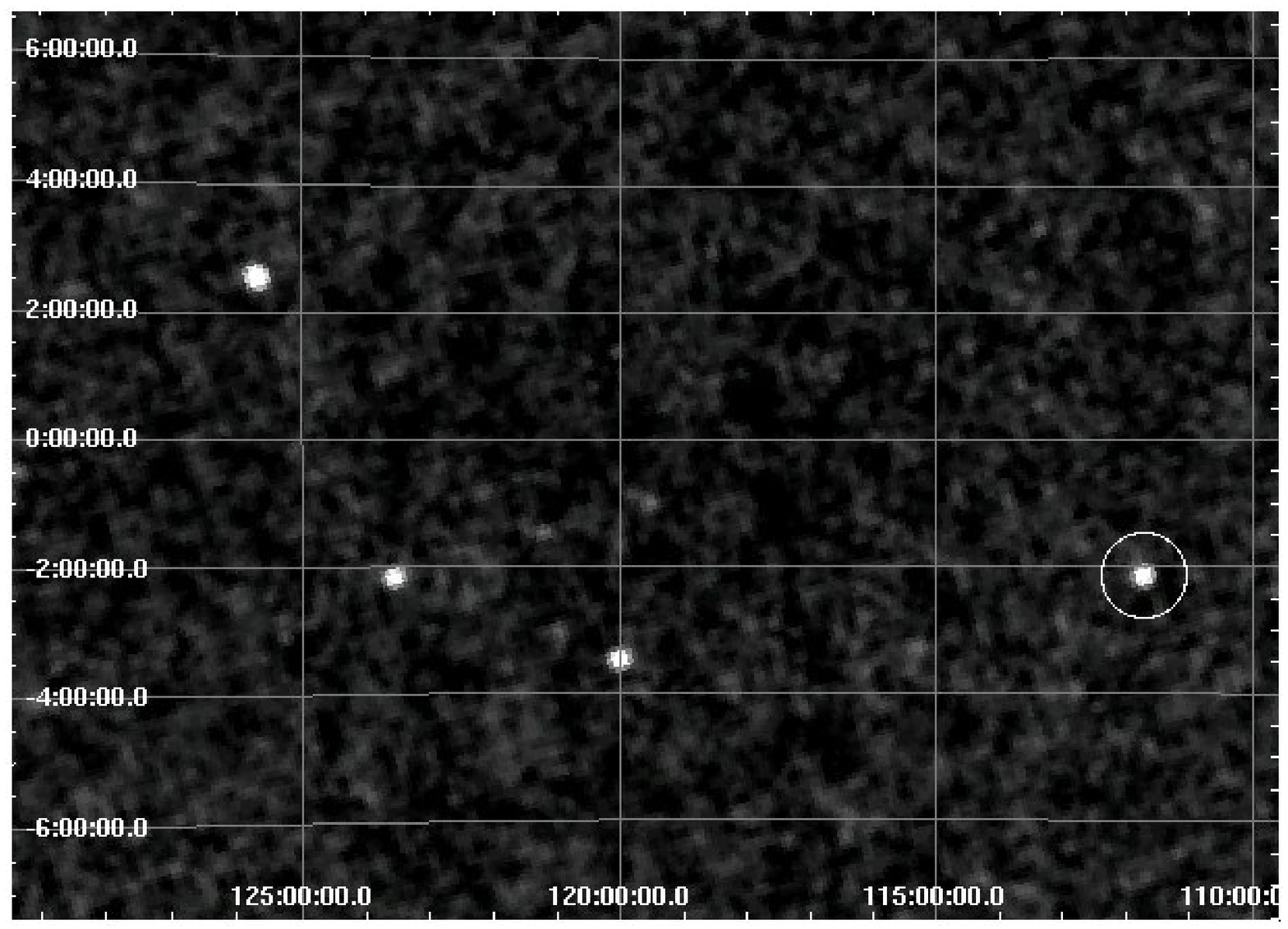,width=0.45\textwidth}
\vbox{
\psfig{figure=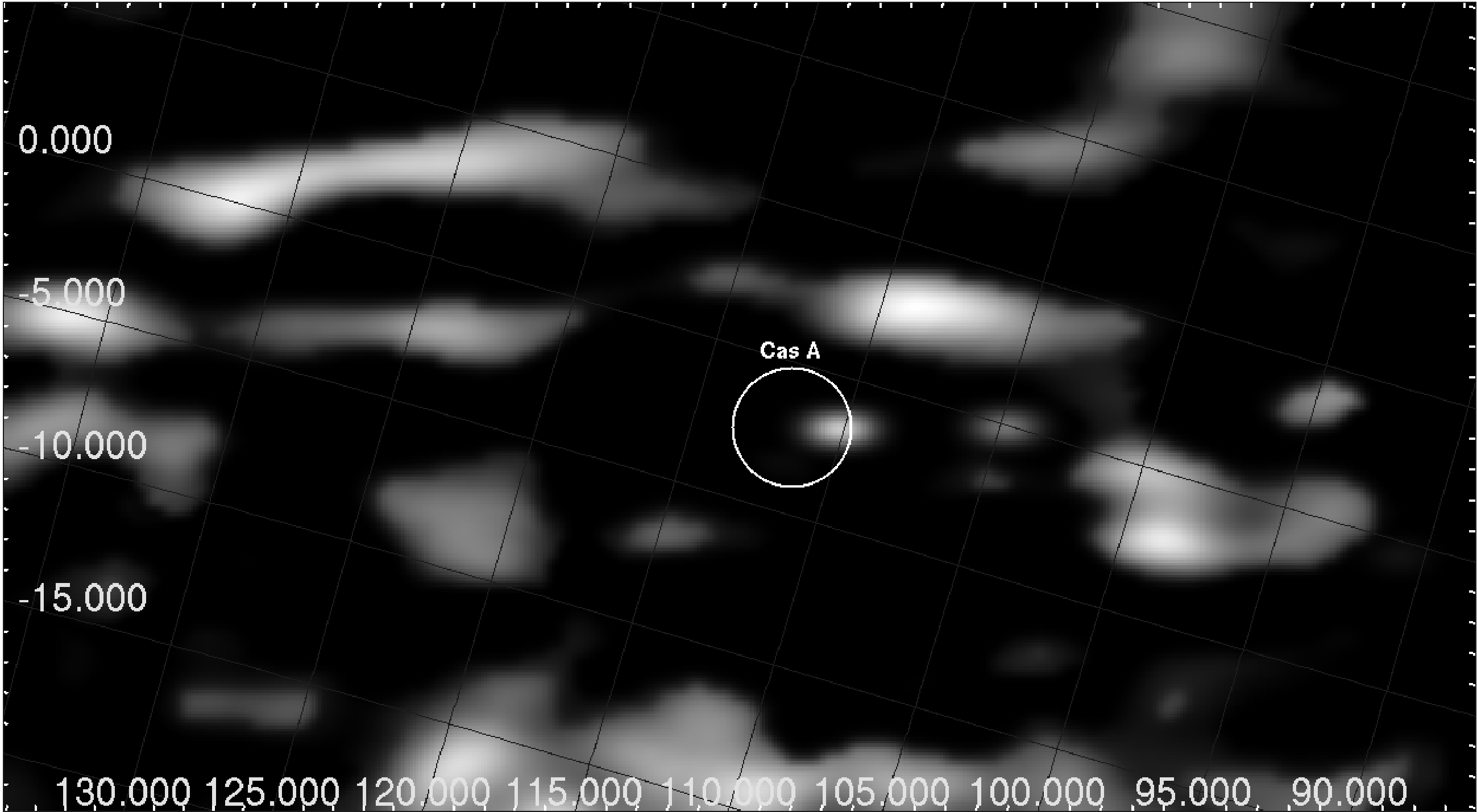,width=0.45\textwidth}
\vskip 5mm
}
}
\centerline{
\psfig{figure=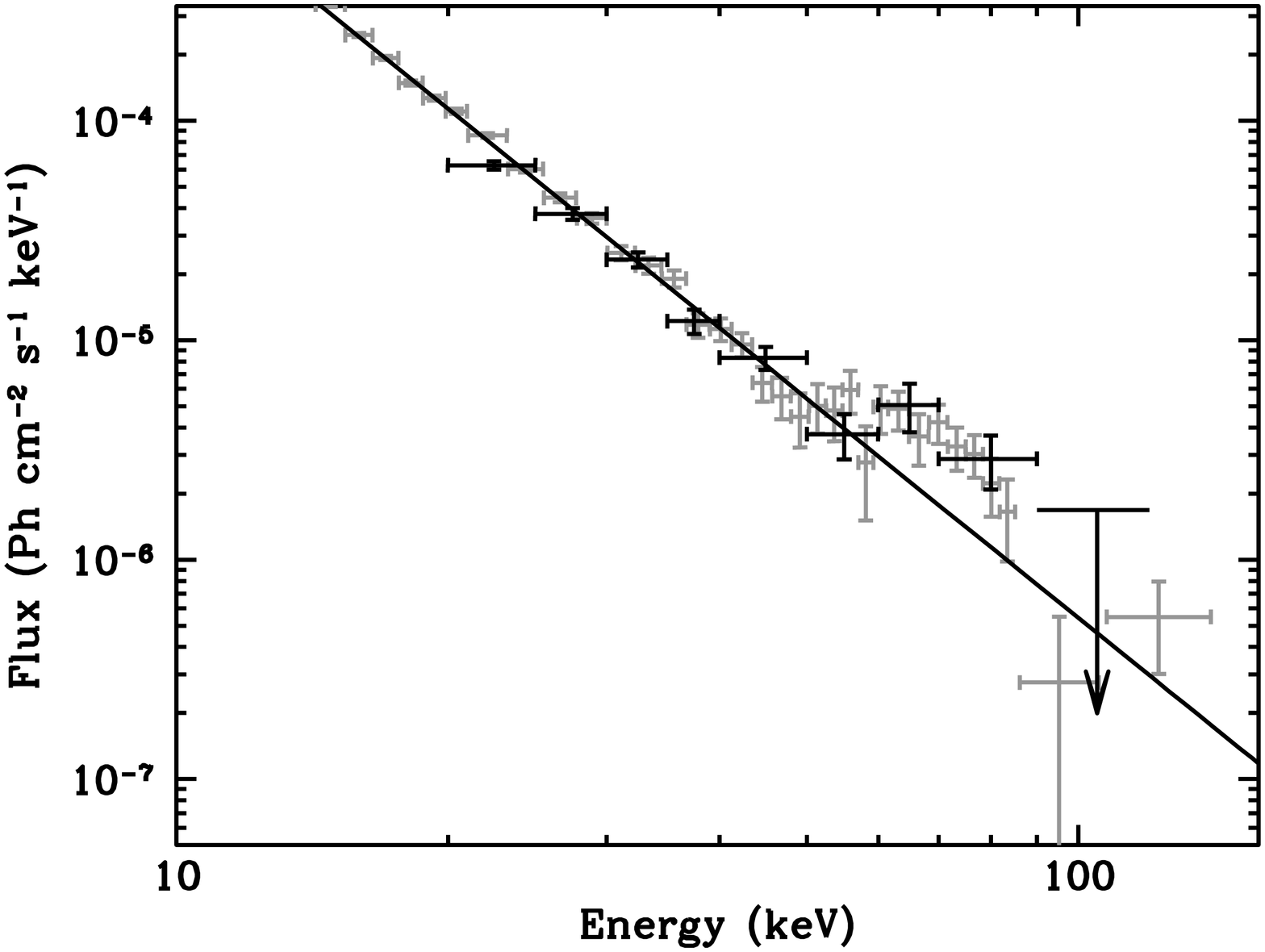,angle=-90,width=0.6\textwidth}
}}

\caption{Left top panel: \integr-ISGRI significance map of the \casa\ field
for the 20-50 keV band. 
The brightest source in the field is the high mass X-ray binary 2S~0114+650
\citep[e.g.][]{hall00}.
Right top panel: \integr-SPI significance map for the 1142-1172 keV band,
selecting both single and double events (SE and ME2).
%A $3\sigma$\ upper limit to 
%The highest significance level is $2.4\sigma$\ corresponding
%to a total flux of $7\times10^{-5}$~\fluxunit.
The location of \casa\ is indicated by a circle with a radius of
2\degr, the coordinates shown are Galactic.  
Bottom panel:
\integr-ISGRI spectrum of \casa. The \sax-PDS points
are shown in light gray for comparison (Fig.~\ref{pdsspec}). 
%The solid line is the
%best fit power law continuum spectrum obtained by \citet{vink01a}.
\label{ibisspec}}
\end{figure*}

\subsection{\casa}

As a known \tiff\ source \casa\ is an important \integr\ target.
The main goal of observing \casa\ with \integr\
is to obtain an independent measurement
of the 67.9~keV, 78.4~keV and 1157~keV line fluxes, and, moreover,
to measure, or put strong limits, on the \tiff\ kinematics.
The kinematics is of particular interest, given the evidence that
the explosion of \casa\ was asymmetric, and some fraction of the Fe-rich
ejecta must have been ejected with a velocities of 4500-7800~\kms
\citep[e.g.][]{vink04a}. This Fe is likely to be the decay product of \nifs,
which was, like \tiff, synthesized in the hottest regions, 
deep inside the supernova.
However, only the shocked Fe can be observed in X-rays, so it is possible
that most Fe, and therefore most \tiff, moves with velocities below 1000~\kms\
and has not yet been heated by the reverse shock.
Note that the \gray\ line emission of \tiff, unlike X-ray line
emission, does not depend on the local conditions 
(except if the \tiff\ has been almost completely ionized,
see section \ref{implications}).

Measuring the \tiff\ velocity may therefore cast new light on the
explosion mechanism of core-collapse supernovae.
Unfortunately, the line sensitivity
of SPI decreases substantially as a function of line broadening.
For example,  for a narrow line one can obtain a $3.4\sigma$\ detection
of the 1157~keV line with SPI with a 2~Ms exposure, but this is only
$1.9\sigma$ 
if the lines are broadened with 2000~\kms\ FWHM.\footnote{This is based
on the \integr\ ``Observation Time Checker'' 
(\url{http://astro.estec.esa.nl/Integral/isoc/operations/html/OTE.html}).}
One can increase the sensitivity, however, by using not only the single 
detector events, but also events that are registered in two detectors,
due to a Compton interaction.

Up to January 2005, \integr\ has completed the first
two cycles of observations of 
\casa, with a total exposure of 3~Ms.\footnote{The \integr\ \casa\ observations
are amalgamated with observations of Tycho's supernova remnant, 
which is in the same field of view and may be a source of \tiff as well,
if it is the result of a sub-Chandrasekhar white dwarf explosion 
\citep{woosley94}.}
The December 2004 observations have not yet been processed at the time
of this writing, but the first 1.6~Ms of \integr\ observations
have resulted in a detection of \casa\ by 
the IBIS-ISGRI instrument.
The data were analyzed
with the standard \integr\ software package {\em OSA} v3 and, more recently,
v4.2. In order to minimize the flux errors due to calibration uncertainties,
we estimated fluxes by comparing the \casa\ source signal per energy bin
directly with that of the Crab nebula. 
For the conversion to flux units we used the parametrization of the
Crab spectrum by \citet{willingale01}.

The IBIS-ISGRI spectrum is shown in Fig.~\ref{ibisspec} and is fully consistent
with the \sax-PDS spectrum. There is a clear signature of excess emission 
with respect to a power-law continuum emission,
and the detection significance of the 67.9~keV line is at the $3\sigma$\ 
level with a flux of $(2.3\pm0.8)\times 10^{-5}$~\fluxunit.
The 78~keV is not detected, but fitting
the 67.9~keV and 78.4~keV jointly, using a power-law continuum,
both
the measured flux per line $(1.2\pm0.6)\times 10^{-5}$~\fluxunit,
and the power law index, $3.27\pm0.15$, are consistent with the
\sax-PDS measurements.

Further results based on the full cycle 1 and 2 data are expected in
the near future (Vink et al. 2005, in preparation). 
In addition, 2.5~Ms of observation time have
been approved for cycle 3, which is
is expected to yield a detection of the continuum above 90~keV,
if the continuum has indeed a power law shape.
This is important, as the apparent low 78.4~keV line flux may in fact
be caused by a steepening of the continuum above $\sim 60$~keV.
This would be of great interest, as a steepening
is predicted for all synchrotron and some bremsstrahlung models 
(section \ref{secCasA}).

It is clear that \integr-IBIS 
has not yet improved upon the \sax\ measurement of
\casa. However, this is expected to change in the near future.
Moreover, IBIS is an imaging instrument, whereas \sax-PDS was
a narrow field instrument using a collimator \citep{frontera97}. As a result
IBIS can observe a large field and separate point sources, but
the use of coded masks requires a larger detector area,
which causes a relatively large background signal. 
The advantage is that we can now resolve several
hard X-ray sources in the field of \casa, which allows us
to exclude the possibility that
a hard X-ray source was contaminating the \sax-PDS spectrum of 
\casa. Moreover, with IBIS it is possible to observe simultaneously
\casa\ and Tycho, and in addition it reveals the presence of many
unresolved sources, two of which were
not previously known as hard X-ray sources 
\citep{denhartog04a,denhartog04b}.

Figure~\ref{ibisspec} shows an IBIS-ISGRI and a SPI significance map.
The SPI map is made with the {\em spiros} imaging software using
the mean count modulation (MCM) method for the background modeling 
\citep{skinner03}.
Both single and double detector events were used, excluding
time intervals with high background rates.
The map shown in  Fig.~\ref{ibisspec} 
covers the energy range 1142-1172~keV,
which corresponds to a total velocity width of $\pm$7800~\kms,
the velocity width that can be expected
if most of the \tiff\ is situated in or is interior to the bright X-ray
shell \citep{willingale02}.
Neither this map, nor the analysis of the available data in
a narrow energy range from 1155 keV to 1159~keV yield a significant
detection. 
However, a 1.8$\sigma$ excess is found within 2\degr\ of the position
of \casa\ with a flux of $1.8\times10^{-5}$~\fluxunit,
similar to the flux measured by \sax\ and \integr-IBIS.

A preliminary analysis shows that 
narrow line \tiff\ emission ($\Delta E=4$~keV,$\Delta v=1000$~\kms) 
can almost be excluded at the 2-3$\sigma$\ level,
as the 3$\sigma$\ upper limit for emission is  
$3.1\times10^{-5}$~\fluxunit\, the flux level found by \comptel\ 
\citep{schoenfelder00}.
and the 2$\sigma$\ upper limit is  $1.7\times10^{-5}$~\fluxunit.
It is, however, prudent to allow for a 10\% systematic
uncertainty in the flux calibration.
Note that without a detection, the exclusion of narrow line emission
ultimately depends on the accuracy with which the overall \tiff\ emission
is known. The planned observations of \casa\ by \integr\ will help
therefore in two ways: On the one hand IBIS-ISGRI will allow us to make
more reliable total flux estimates, on the other hand SPI
will provide either more stringent upper limits, which will help
to constrain the line broadening, or a detection, in which case
one can actually measure the line broadening.

\section{Concluding remarks}
Gamma-ray line emission from explosive nucleosynthesis products provides
a view of what happens deep inside supernova explosions.
Explosion asymmetries and turbulence during the explosion can be deduced
from the time of emergence, and Doppler broadening of line emission
from the
short lived radioactive elements \nifs\ and \cofs.
The longer lived radioactive element \tiff\ 
can be used to probe the explosion within
a few hundred years after the supernova event.
Although, \gray\ astronomy is a specialized field,
its results are of general astrophysical importance, as the nature
of both Type Ia and core-collapse supernovae are not well understood.
This is even more important now that Type Ia supernovae are used
to probe the history and fate of the universe, and gamma-ray bursts appear
to be connected with core-collapse supernovae.

I have shown that \gray-astronomy has already fulfilled part of its promise.
The detection of \cofs\ lines from SN 1987A 
has provided us with clues about the mixing of supernova ejecta, although the
theoretical understanding of the observations is not yet complete.
The detection of \tiff\ from \casa\ by \comptel\ and \sax, 
on the other hand, suggests 
that theoretical models underpredict the \tiff\ yield.

The work that started with \smm, \comptel\ and \sax-PDS\ and other experiments
is continued with ESA's \integr\ mission.
For the more distant future several satellite missions are envisioned
ranging from focusing telescopes, using grazing incidence telescopes
with multi-layer coating, 
von Laue lenses \citep[i.e. using Bragg-crystals ][]{vonballmoos04}, or
a new Compton telescope \citep{kurfess04}.

At the time of this writing \integr\ is in its second observing cycle,
having already produced the first results on 511~keV, $^{26}$Al, and \tiff\
\gray\ line emission, which will be improved upon in the near future.
And with some luck a not too distant Type Ia supernova may produce detectable
\gray\ line emission, or, we might even be as lucky
as Tycho Brahe and Johannes Kepler and observe a Galactic supernova.

\section*{Acknowledgments}
I thank Peter den Hartog,  Hans Bloemen and J\"urgen Kn\"odlseder for 
their help in analyzing the \integr\ data, and Matthieu Renaud and Francois
Lebrun for discussing \integr-IBIS data analysis.
I am grateful to Roland Diehl and Cara Rakowski 
for their careful reading of the manuscript.
This work has profited from discussions during two
meetings on supernova remnants at the International Space Science
Institute (ISSI) in Bern.

%\bibliographystyle{cospar}
%{\small
%\bibliography{apj-jour,snrs,gamma,sn}
%}

\end{document}